\documentclass[aps, prl, reprint, superscriptaddress]{revtex4-2}

\draft % marks overfull lines with a black rule on the right
\usepackage{graphicx}
\usepackage{dcolumn}
\usepackage{bm}
\usepackage{longtable}
\usepackage[colorlinks=true, allcolors=blue]{hyperref}
\usepackage{amsmath}

\begin{document}

%\preprint{}

\title{Self-focusing and self-compression of intense pulses via ionization-induced spatiotemporal reshaping} %Title of paper



\author{Xiaohui Gao}
\email[]{gaoxh@utexas.edu}
%\thanks{}
\affiliation{Department of Physics, Shaoxing University, Shaoxing, Zhejiang 312000, China}
\author{Bonggu Shim}
\affiliation{Department of Physics, Applied Physics and Astronomy, Binghamton University, Binghamton, New York 13902, USA}

\date{\today}

\begin{abstract}
Ionization is a fundamental process in intense laser-matter interactions, and is known to cause plasma defocusing and intensity clamping. Here, we investigate theoretically the propagation dynamics of an intense laser pulse in a helium gas jet in the ionization saturation regime, and we find that the pulse undergoes self-focusing and self-compression through ionization-induced reshaping, resulting in a manyfold increase in the laser intensity. This unconventional behavior is associated with the spatiotemporal frequency variation mediated by ionization and spatiotempral coupling. Our results illustrate a new regime of pulse propagation and open up an optics-less approach for raising the laser intensity.
\end{abstract}

\maketitle %\maketitle must follow title, authors, abstract

%\section{}
%\label{}
%\subsection{}
%\subsubsection{}
The nonlinear propagation of intense pulses in a transparent medium has attracted great attention because it is remarkably rich in its physics and practically important for various applications such as high harmonic generation~\cite{Popmintchev2015S} and plasma waveguide formation~\cite{Shalloo2018PRE}. In many applications it is desirable that the pulse undergoes self-focusing and self-compression, such that the light is compressed in space and in time and propagates at a high intensity. 
These self-actions are primarily found in the weak ionization regime~\cite{Couairon2007PR, Champeaux2003PRE, Ward2003PRL,Mysyrowicz2008NJP} or the relativistic intensity regime~\cite{Monot1995PRL, Tsung2002PNASU}, where the intensity dependent Kerr refractive index or the relativistic dependence of electron mass on its quiver velocity can be exploited. 

For an intensity in the range of $10^{15}$ - $10^{17}\,$W/cm$^2$, Kerr nonlinearity is depleted by ionization and relativistic nonlinearity is absent,  making it elusive to observe self-focusing. While the ionization nonlinearity also plays an important role in the propagation, the primary consequence is defocusing. The refractive index of the ionized gas medium is 
\begin{equation}
n=\sqrt{1-\omega_p^2/\omega^2}, 
\label{index}
\end{equation}
where $\omega$ is the angular frequency of the pulse, $\omega_p=\sqrt{\rho_ee^2/\epsilon_0m_e}$ is the plasma frequency, $\rho_e$ is the electron density, $e$ is the elementary charge, $\epsilon_0$ is the permittivity of the vacuum, and $m_e$ is the electron mass. Thus the higher the laser intensity is, the greater the plasma density accumulates and the lower the refractive index becomes.  This reduces the refractive index on the beam axis and leads to the well-known plasma defocusing. Refocusing occurs only under rare conditions, such as the use of tightly focused pulses to produce a ring-shaped plasma~\cite{Gao2019OLb} or the use of a clustered gas to create overdense plasmas~\cite{Alexeev2003PRL}. Self-focusing or other approach such as flying focus~\cite{Froula2018NP} and self-generated plasma capillary guiding~\cite{Tosa2003PRA} is required to mitigate plasmas defocusing. While ionization leads to refraction in the spatial domain, it causes blue-shift~\cite{Wilks1988PRL} and hence spectral broadening in the spectral domain. Combined with appropriate self-induced chirp compensation, this can lead to  self-compression. This effect is initially demonstrated numerically in one dimensional case~\cite{Kim1990PRA} with ionization-induced chirp~\cite{Mysyrowicz2008NJP} and is weaker when defocusing is considered~\cite{Sergeev1992PRA}. Ionization-induced compression was also demonstrated experimentally in capillaries or under tight focusing, where the required chirp is mediated by the strong spatiotemporal coupling~\cite{Wagner2004PRL, He2014PRL}. 

In this Letter, we model the propagation dynamics of intense femtosecond pulses above ionization saturation intensity in a thin helium gas jet, and we show a novel regime of pulse propagation in highly ionized plasmas with simultaneous self-focusing and self-compression, which is in strong contrast to the plasma defocusing. This new self-focusing mechanism occurs at an intensity regime well above that of Kerr-induced self-focusing and well below relativistic intensity, and is attributed to an index profile corresponding to a positive lens as a result of the transverse variation of instantaneous frequency. 

We adopt the nonlinear envelope equation in the spectral domain~\cite{Couairon2011EPJST} in cylindrical coordinates since radial symmetry is maintained in typical laser-gas jet interaction~\cite{Chessa1999PRL}. The pulse propagation equation is given by
\begin{equation}
\frac{\partial \mathcal{\tilde E}}{\partial z}=\frac{i\,\Delta_\bot\mathcal{\tilde E}}{2k(\omega)}+i[k(\omega)-\kappa(\omega)]\mathcal{\tilde E} +\frac{i}{2k(\omega)}\frac{\omega^2}{\epsilon_0c^2}\mathcal({\tilde P}_{NL}+i \frac{\mathcal{\tilde J}}{\omega}).
\label{prop}
\end{equation}
Here $\mathcal{\tilde E}$ denotes the Fourier transformed complex envelope of the electric field, $z$ is the distance, $k(\omega)=\omega n(\omega)/c$ is the wavenumber,  $n(\omega)$ is the refractive index of the medium, $c$ is the vacuum speed of light, $\kappa(\omega) \equiv k_0+(\omega-\omega_0)/v_g$, $k_0$ and $\omega_0$ is the wavenumber and the angular frequency of the carrier wave, $v_g$ denotes a constant velocity of reference frame with respect to the laboratory frame, which we take equal to the group velocity in the neutral medium,
$\mathcal{\tilde P}_{NL}$ and $\mathcal{\tilde J}$ denote the Fourier transformed nonlinear polarization and current, which are given by
\begin{eqnarray}
\mathcal {\tilde P}_{NL}&=&2\epsilon_0n_0\mathcal{FT}[{n_2^{\text{eff}}I\mathcal E}],\\
\mathcal{\tilde J}&=&-\frac{1}{2}\mathcal{FT}[\sum_jW_jU_j{\rho_j{ I}^{-1}\mathcal{E}}]+\frac{e^2}{m_e}\frac{i}{\omega}{\mathcal{FT}[ \rho_e \mathcal{E}}].
\end{eqnarray}
Here $n_2^\text{eff}=\sum_j n_{2,j}^\text{STP}\rho_j/N_A$ is the effective nonlinear refractive index, $n_{2,j}^\text{STP}$ is the nonlinear refractive index of ions in charge state $j$ at standard temperature and pressure (STP), $\rho_j$ is the density of ions in charge state $j$, $N_A$ denotes gas density at STP, $I$ is the laser intensity, $W_j$ and $U_j$ are the ionization rate and ionization potential of ions in charge state $j$, respectively. The medium response is calculated in the time domain and $\mathcal{FT}$ represents Fourier transform. The ion density $\rho_i$ is calculated using the rate equations given by
\begin{eqnarray}
\frac{d\rho_{0}}{dt}&=&-W_0\rho_{0},\\
\frac{d\rho_{j}}{dt}&=&-W_j\rho_{j}+W_{j-1}\rho_{j-1}, \text{for } j>1, 
\end{eqnarray}
and $\rho_e=\sum_j \rho_j$. Note that quantities of neutral atoms are denoted by $j=0$. We have ignored collision processes including electron-neutral collision, impact ionization, and recombination because the gas pressure is low.

In all simulations, we assume an unchirped 800-nm Gaussian input pulse in space and time. The full-width at half maximum (FWHM) pulse duration is 50 fs, and the beam waist $w_0$ is $40\,\mu$m. The focus is at the entrance of the gas jet. The energy is varied to get different initial intensities. We assume a gas of 100-mbar pressure with a uniform gas density and a length of 2.5-mm. The linear index of the gas is modeled using the Sellmeier relation. The nonlinear refractive index of neutral atom and ions at standard temperature and pressure (STP) are taken from theoretical calculation~\cite{Tarazkar2016PRA}, which are $n_{2,0}^\text{STP}=0.348\times10^{-20}$cm$^2$/W and $n_{2,1}^\text{STP}=0.031\times10^{-20}$cm$^2$/W, and the ionization rate is calculated using ADK formula~\cite{ Ammosov1986SPJ}.

\begin{figure}[htbp]
\centering
\includegraphics{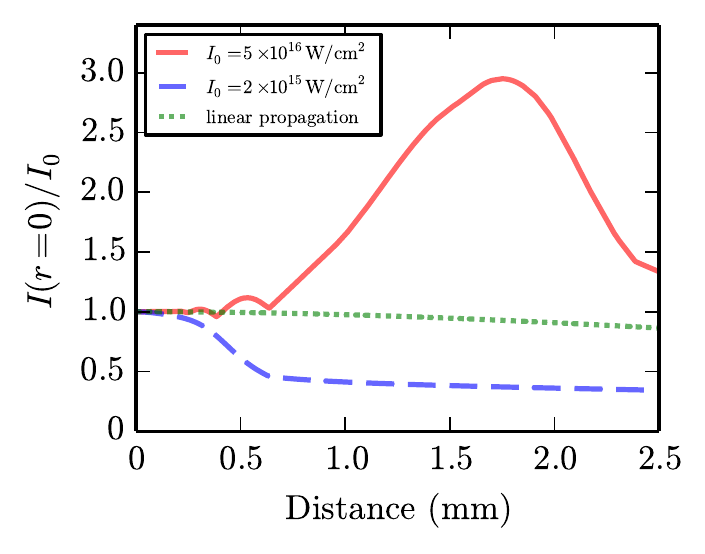}
\caption{Enhancement factor of the peak intensity vs propagation distance. The initial intensities of the solid red and dashed blue lines are $5\times10^{16}\,$W/cm$^2$ and $2\times10^{15}$\,W/cm$^2$, respectively. The dotted green line represents the intensity evolution of linear propagation.}
\label{fig1} 
\end{figure}%
Figure~\ref{fig1} plots the peak intensity as a function of distance for different initial intensities. The dashed blue curve shows the evolution of on-axis peak intensity at an initial intensity of $2\times 10^{15}$ W/cm$^2$. The peak intensity drops quickly as expected due to plasma defocusing. However, for an initial intensity of $5\times 10^{16}$ W/cm$^2$, we observe a drastic intensity increase instead of plasma defocusing. As shown by the solid red curve, after a relative stable propagation of 0.7 mm, the intensity grows rapidly with the peak almost tripled in just one millimeter.  Here $P/P_\text{cr}=0.45$.
The Kerr effect of neutral atoms or ions plays little role in this dynamics, because switching off $n_2$ of atoms or ions does not alter this behavior. Similar behavior is also observed in simulations using carrier-resolved model gUPPElab~\cite{Kolesik2004PRE,Andreasen2012PRE}. 

\begin{figure}[htbp]
\centering
\includegraphics{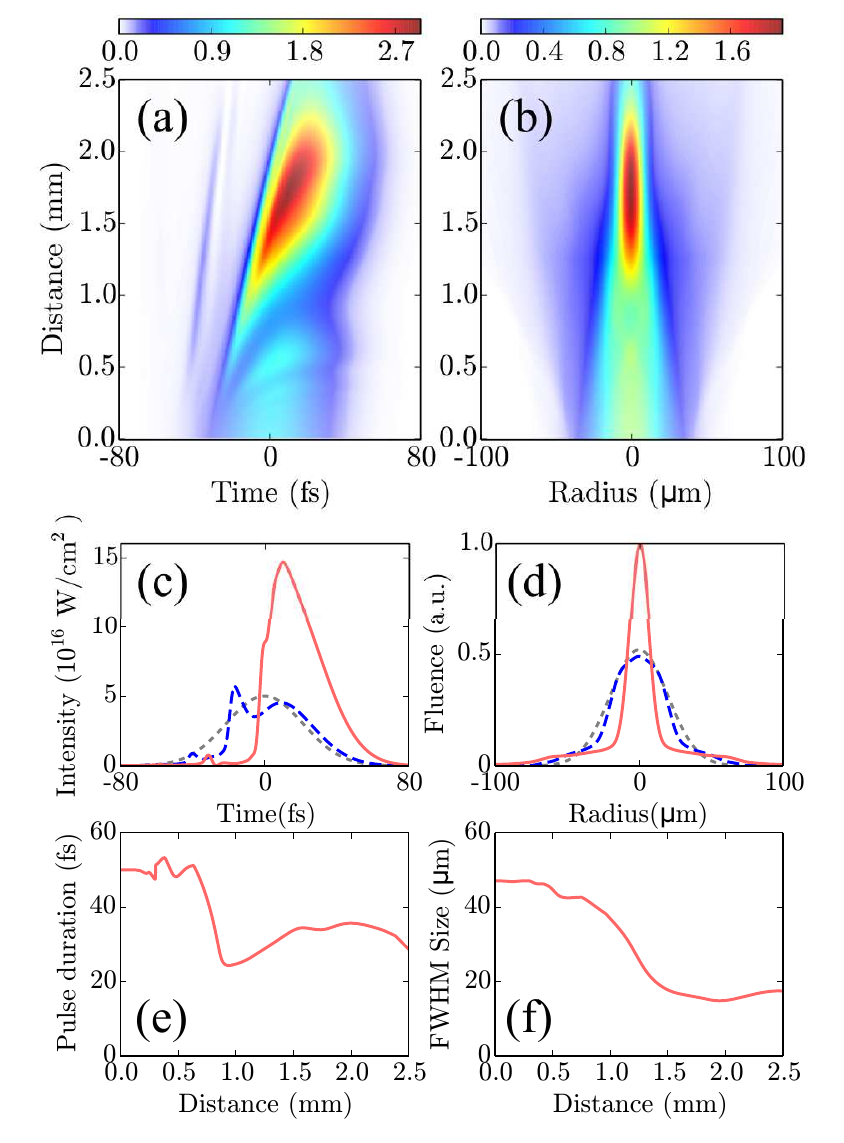}
\caption{(a) Evolution of the on-axis ($r=0$) temporal profiles during the propagation. The profile is normalized by the peak intensity at $z=0$.  (b) Evolution of the fluence distribution normalized by the initial peak fluence during the propagation (c) Temporal profiles at $z=0$ (short-dashed curve), $0.7$\,mm (dashed curve), and $1.7$\,mm (solid curve). (d) Fluence distribution at the same three locations. (e) On-axis pulse duration vs. propagation distance. (f) FWHM beam size vs. propagation distance. }
\label{fig2} 
\end{figure}%
To elucidate whether the intensity enhancement is due to self-focusing or self-compression or both, we present the evolution of on-axis temporal profile and fluence distribution for $I=5\times 10^{16}$ W/cm$^2$ in Fig.~\ref{fig2}(a) and \ref{fig2}(b), respectively. The lineouts at three locations $z=0$, $0.7$\,mm, and $1.7$\,mm are given in Fig.~\ref{fig2}(c) and \ref{fig2}(d). The pulse duration and the FWHM size as a function of distance are plotted in Fig.~\ref{fig2}(e) and \ref{fig2}(f), respectively. These results evidently show self-compression and self-focusing.  The group velocity in plasmas is smaller than that in the neutral medium. Thus overall the power is shifted backward, as shown in Fig.~\ref{fig2}(a). Figure~\ref{fig2}(c) shows a shock formation at the leading edge of the pulse. This contrasts with that in Kerr-induced self-focusing, which emerges at the trailing edge of the pulse. Figures~\ref{fig2}(b), \ref{fig2}(d), and ~\ref{fig2}(f) show that in the transverse dimension, the light in the periphery diffracts, while that in the central region contracts, reducing the beam size by more than half. 

\begin{figure}[htbp]
\centering
\includegraphics{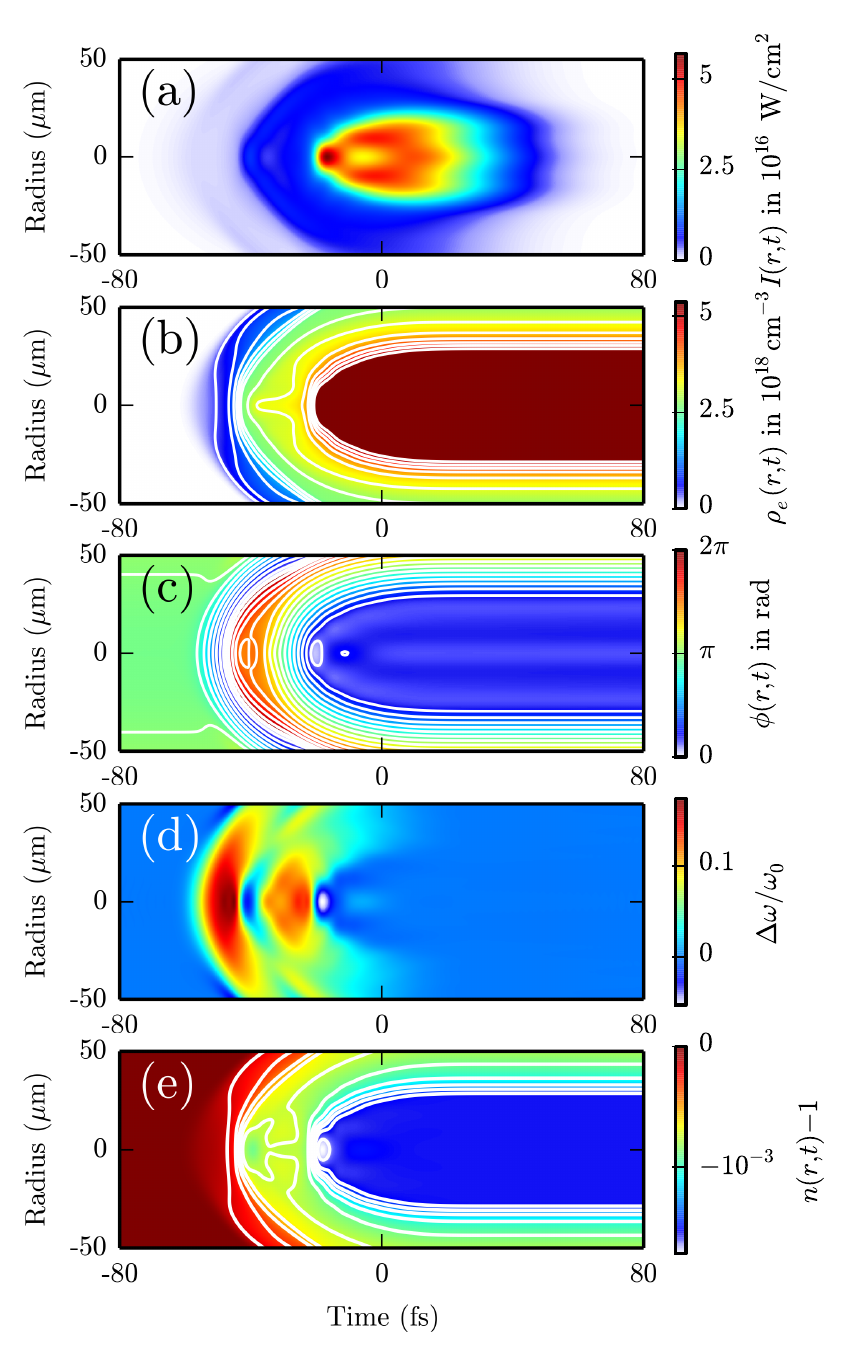}
\caption{Spatio-temporal profiles of the following quantities at $z=0.7$ mm (a) intensity $I(r,t)$,  (b) electron density
$\rho_e(r,t)$, (c) phase $\phi(r,t)$, (d) frequency shift $\Delta\omega(r,t)/\omega_0$, (e) refractive index change $n(r,t)-1$.  The solid white lines in panel (b), (c) and (e) are contour plots.}
\label{phase} 
\end{figure}%
Figure~\ref{phase} shows the profiles of (a) intensity $I(r,t)$, (b) electron density $\rho_e(r,t)$, (c) phase $\phi(r,t)$, (d) the corresponding frequency shift $\Delta\omega(r,t)/\omega_0$, and (e) the refractive index change $n(r,t)-1$ for $I=5\times 10^{16}$ W/cm$^2$ at $z=0.7$\,mm.   The electron density in Fig.~\ref{phase}(b) shows a weak off-axis maximum at around $t\approx-50\,$fs. While a ring-shaped plasma can lead to beam focusing~\cite{Gao2019OLb, Gao2019OLa}, this is unlikely the cause here because it appears only briefly in the leading edge and affects only a small fraction of the energy. The frequency shift shown in Fig.~\ref{phase}(d) is calculated using $\Delta\omega(r,t)=d\phi(r,t)/d t$. Two belts of strong blueshift basically correspond to the ionization to He$^+$ and He$^{2+}$, respectively. The refractive index is calculated using Eq.~\ref{index} and the distribution is shown in Fig.~\ref{phase}(e). We do not observe noticeable difference  when we set  $n_2=0$. 

\begin{figure}[htbp]
\centering
\includegraphics{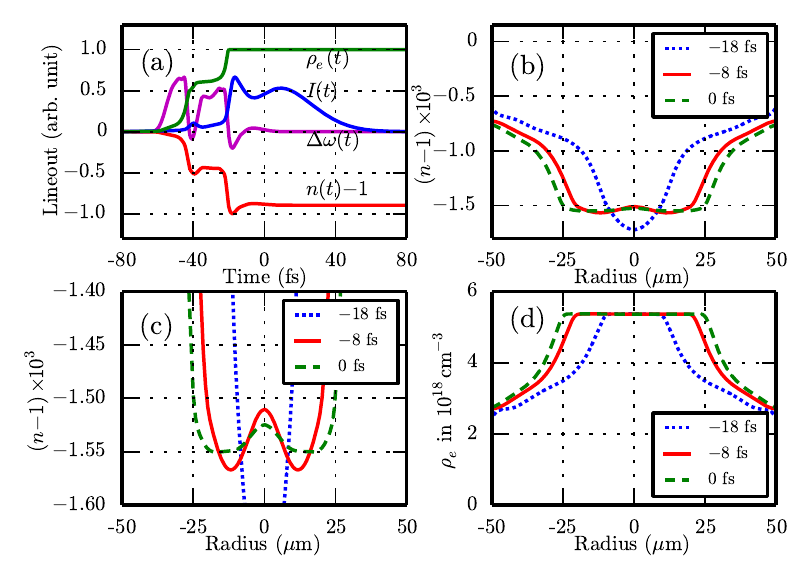}
\caption{(a) Time evolution of electron density, intensity, frequency variation, and refractive index on the beam axis ($r=0$) at $z=0.7$\,mm. (b) Transverse distribution of refractive index change $n(r)-1$ for three time slices at $z=0.7$ mm. (c) Vertical zoomed view of panel (b). (d) Transverse distribution of $\rho_e$ for three time slices at $z=0.7$ mm.} 
\label{lineout}
\end{figure}%
For a better illustration, the time evolution of the refractive index change  and other quantities on the laser axis ($r=0$) is shown in Fig.~\ref{lineout}(a), and the transverse pattern of several time slices of $n(r,t)$ is shown in Fig.~\ref{lineout}(b) and Fig.~\ref{lineout}(c). At $z=0.7$\,mm, $\rho_e$ increases monotonically but the refractive index does not decrease monotonically. The rise of the refractive index is due to the increase of the instantaneous frequency while the increase of electron density is minimal or saturated. At $z=0.7$\,mm, $d\rho_e/dt$ does not yield the same profile as $\Delta\omega$. This is because the frequency shift is an accumulated effect of the propagation, during which the time profile of the intensity and plasma are evolving. Moreover, spectral components may be redistributed due to group velocity dispersion. Indeed, the Gouy phase is frequency dependent~\cite{Porras2003PRE}. Gouy phase stems from transverse spatial confinement~\cite{Feng2001OL}. The blue-shifted frequency component have a smaller transverse spatial size. This affects group velocity dispersion and causes spatiotemporal coupling. 
In a plasma, the phase velocity is $v_p=c/n$ and the group velocity is $v_g=cn$. When the index increases with the time, the group velocity also increases. This shifts the power forward within the pulse, causing self-steepening of the leading edge. 
In the transverse dimension, there is a time period around $t=0$ where the refractive index $n(r)$ shows an on-axis local maximum while $\rho_e$ is saturated, as shown in Fig.~\ref{lineout}(c) and Fig.~\ref{lineout}(d). The intensity at that time remains high. Thus a large portion of the pulse experiences a positive lensing effect. Considering a gradient-index lens with a parabolic index profile, the focal length is~\cite{Riedl2001}
\begin{equation}
f=\frac{R}{\sqrt{2n_0\Delta n}},
\label{eq7}
\end{equation}
where $R$ is the lens radius, $n_0$ is the index at the center, $\Delta n$ is the index difference between the lens center and edge.  
If we assume a constant parabolic index profile with 
$R=12\,\mu$m and $\Delta n=5\times10^{-5}$, we have $f=1.2\,$mm.
This estimation is consistent with the simulation where the beam size is reduced approximately by half in one millimeter. Note that the index profile is evolving during the propagation, Estimate from Eq.~\ref{eq7} is reasonable only when this parabolic index profile is maintained with a length comparable to the focal length. At lower intensity, the parabolic index profile may appear for a shorter propagation distance thus the enhancement gets weaker and ends earlier.

\begin{figure}[!ht]
\centering
\includegraphics{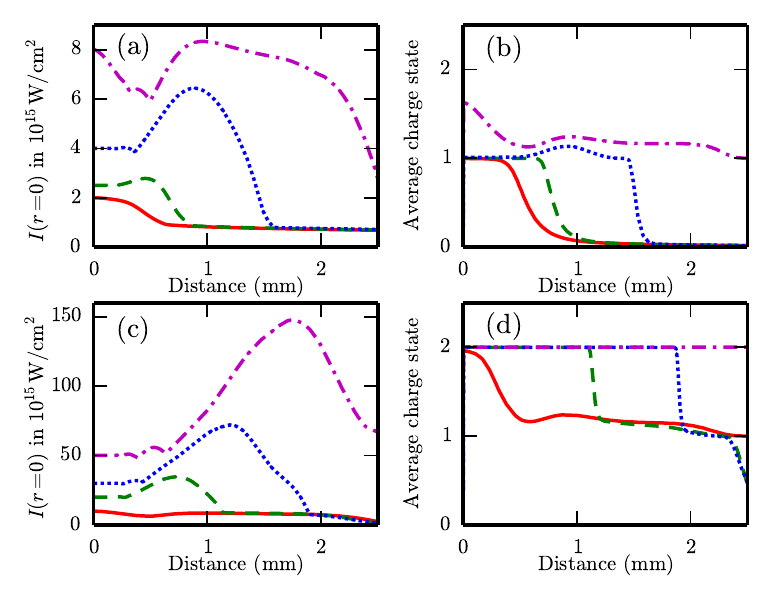}
\caption{(a) Peak intensity vs. distance for input intensities at $2\times10^{15}$ (solid red),  $2.5\times10^{15}$ (dashed green),  $4\times10^{15}$ (dotted blue), and $8\times10^{15}$ W/cm$^2$ (dash-dotted magenta). (b) Average charge state vs. distance for the corresponding input intensities. (c) Peak intensity vs. distance for input intensities at $1\times10^{16}$ (solid red),  $2\times10^{16}$ (dashed green),  $3\times10^{16}$ (dotted blue), and $5\times10^{16}$ W/cm$^2$ (dash-dotted magenta). (d) Average charge state vs. distance for the corresponding input intensities. } 
\label{fig5}
\end{figure}%
This intensity enhancement can be observed over a wide range of intensities. Figure~\ref{fig5} shows the evolution of intensity and average charge state for different intensities. A moderate increase of the input intensity from $2\times10^{15}\,$W/cm$^2$ to $2.5\times10^{15}\,$W/cm$^2$ changes the dynamics from plasma defocusing to intensity enhancement. At $4\times10^{15}\,$W/cm$^2$,  the enhancement is substantial due to the intensity gap where ionization saturates at $Z=1$. 
When the intensity increases to $8\times10^{15}$ W/cm$^2$, the pulse undergoes plasma defocusing for the first half millimeter due to second ionization of helium. When the input intensity is above saturation intensity of ionization to He$^{2+}$, the intensity enhancement occurs again as shown in Fig.~\ref{fig5}(c). Here we observe that as the intensity increases, the enhancement is stronger and its peaks are located at farther distances. 
We also find that increasing the pressure affects the enhanced peak intensity minimally, but the location of the peak is roughly inversely proportional to the pressure. In our simulations, we do not observe more than a single-cycle of focusing when we use a longer jet. Previous simulations have shown that the pulse intensity increases in helium gas~\cite{Gonzalez2014JPBAMOP}. Because the gas pressure is 26 bar and the intensity is below the saturation intensity, the self-focusing and self-compression observed in their simulations are due to Kerr effect rather than ionization.

In conclusion, we have demonstrated a novel regime of intense pulse propagation in gas jets through numerical investigation. For pulses above the ionization saturation intensity, we find that self-focusing and self-compression occur simultaneously, giving rise to an impressive increase in the laser intensity. The transverse variation of instantaneous frequency gives rise to an index profile of a positive lens around the central time slice of the pulse, enabling efficient self-focusing. These results can be relevant for high-field applications such as high-order harmonic generation in plasmas~\cite{Popmintchev2015S, Gao2018OL}.   

\noindent{\bf{Funding.}} Natural Science Foundation of Zhejiang Province (LY19A040005); National Science Foundation (NSF) (PHY-1707237).

% If in two-column mode, this environment will change to single-column format so that long equations can be displayed. 
%\begin{widetext}
%$$\mbox{put long equation here}$$
%\end{widetext}

% Figures should be put into the text as floats. 
% Use the graphics or graphicx packages (distributed with LaTeX2e).
% Put the label that you will use with \ref{} command in the braces of the \label{} command.
%
% \begin{figure}
% \end{figure}

% \begin{table}
% \end{tabular}
% \end{table}

%\begin{acknowledgments}
% Put your acknowledgments here.
%\end{acknowledgments}

\bibliographystyle{aipnum4-2}
%aipnum4-2.bst 2019-01-14 (MD) hand-edited version of apsrev4-1.bst

\end{document}